# Conjectural Equilibrium in Water-filling Games

Yi Su[*] and Mihaela van der Schaar

ABSTRACT

This paper considers a non-cooperative game in which competing users sharing a frequency-selective interference channel selfishly optimize their power allocation in order to improve their achievable rates. Previously, it was shown that a user having the knowledge of its opponents' channel state information can make foresighted decisions and substantially improve its performance compared with the case in which it deploys the conventional iterative water-filling algorithm, which does not exploit such knowledge. This paper discusses how a foresighted user can acquire this knowledge by modeling its experienced interference as a function of its own power allocation. To characterize the outcome of the multi-user interaction, the conjectural equilibrium is introduced, and the existence of this equilibrium for the investigated water-filling game is proved. Interestingly, both the Nash equilibrium and the Stackelberg equilibrium are shown to be special cases of the generalization of conjectural equilibrium. We develop practical algorithms to form accurate beliefs and search desirable power allocation strategies. Numerical simulations indicate that a foresighted user without any a priori knowledge of its competitors' private information can effectively learn the required information, and induce the entire system to an operating point that improves both its own achievable rate as well as the rates of the other participants in the water-filling game.

## I. INTRODUCTION

Multi-user communication systems represent competitive environments, where devices built according to different standards and architectures compete for the limited available resources. These devices can differ

The authors are with the Dept. of Electrical Engineering (EE), University of California Los Angeles (UCLA), 66-147E Engineering IV Building, 420 Westwood Plaza, Los Angeles, CA, 90095, USA, Phone: +1-310-825-5843, Fax: +1-310-206-4685, E-mail: {yisu, mihaela}@ee.ucla.edu

[*]Corresponding author. Address and contact information: see above.

greatly in terms of their channel conditions, user-defined utilities, action strategies, ability to sense the environment and gather information about competing users, and reason about the available information. Spectrum sharing among multiple competing devices in the interference-limited communication systems provides such a typical scenario. In particular, the performance of each device depends on not only its own power allocation strategy, but also that of the other devices. Individual devices may differ in both their knowledge of the system-wide channel state information (which is for instance constrained by their spectrum sensing abilities and/or information exchange overheads) and their decision making mechanism for choosing their optimal power allocation (which is determined by the implemented protocol).

This paper focuses on the multi-user interaction in frequency-selective Gaussian interference channels. To model the competitive interaction among users, the existing literature investigating this problem often adopts a game theoretic optimization perspective [1]-[12]. Throughout this paper, we focus on a simple yet practical approach that minimizes the complexity of transceivers by treating interference as additive noise. From a particular user's perspective, it is well-known that, for fixed interference power, the optimal power allocation is the so-called water-filling solution. Therefore, the spectrum sharing problem can also be regarded as a *water-filling game*. Specifically, the participants in the water-filling game are modeled as players with individual goals and strategies. They compete or cooperate with each other until they can agree on an acceptable resource allocation outcome. Existing research can be categorized into two types, *non-cooperative* games and *cooperative* games.

First, the formulation of the multi-user environment as a non-cooperative game has appeared in several recent works [1]-[8]. An iterative water-filling (IW) algorithm has been proposed to mitigate the mutual interference and optimize the performance without the need for a central controller [1]. At every decision stage, selfish users deploying this algorithm myopically maximize their achievable rates by water-filling across the entire frequency band until a Nash equilibrium (NE) is reached. Sufficient conditions under which the iterative water-filling algorithm converges to a unique NE are derived and the closed form solution to the water-filling problem is investigated for some special scenarios [2]-[5]. Alternatively, self-enforcing protocols are studied in the repeated game setting, where efficient, fair, and incentive compatible spectrum sharing is shown to be possible by imposing punishment in the case of misbehavior and enforcing users to cooperate [6].

On the other hand, because the IW algorithm may lead to Pareto-inefficient solutions [7], i.e. selfishness is detrimental in the interference channel, there also have been a number of related works studying spectrum sharing in the setting of cooperative games [9]-[12]. Several (near-) optimal algorithms were proposed to address the problem of maximization of a certain user's achievable rate while satisfying the minimum rate requirements of the other competing users. These works assume that users agree to cooperatively maximize a common objective function and require explicit information exchanges among the users.

In short, most of the existing research mainly concentrates on studying the existence and performance of NE in non-cooperative games and developing efficient algorithms to approach the Pareto boundary in cooperative games. Our focus in this paper is on the non-cooperative setting, which explicitly considers the self-interested and competitive nature of individual players. However, most of the existing works in the non-collaborative setting often neglect an important intrinsic dimension of the information-decentralized multi-user interaction. Prior non-cooperative approaches often assume homogeneous users with only the knowledge of their own private information and do not consider users' ability to improve their performance by acquiring and exploring the information of the opponents. The best response strategy of a selfish user that knows its myopic opponents' private information, including their channel state information and power constraints, was first investigated in [8] using the Stackelberg equilibrium (SE) formulation. It was shown in [8] that surprisingly, a foresighted user playing the SE can improve both its performance as well as the performance of all the other users. These results highlight the significance of information availability in water-filling games. However, one key question remains unsolved: how should a foresighted user acquire its desired information and adapt its response?

As opposed to our previous approach, which assumes a foresighted user with perfect knowledge of its competitors' private information [8], we discuss in this paper how the foresighted user without any such a priori knowledge can accumulate this knowledge and improve its performance in water-filling games. We propose that the foresighted user can explicitly model its competitors' response as a function of its power allocation by repeatedly interacting with the environment and observing the resulting interference. The concept of conjectural equilibrium (CE) is introduced to characterize the strategic behavior of a user that models the response of its myopic competing users, and the existence of this equilibrium in the water-filling game is proved. Some previously adopted solutions, including NE and SE, are shown to be special cases of the CE. Practical algorithms are developed to form accurate beliefs and search desirable power allocation strategy. It is shown that, a foresighted user without any a priori knowledge can effectively learn its desired information and guide the system to an operating point having comparable performance to the algorithm in [8], where perfect a priori knowledge is assumed. More importantly, as opposed to the two-user algorithm in [8], the proposed algorithm in this paper can be applied in general scenarios where more than two users exist.

The basic notion of CE was first proposed by Hahn in the context of a market model [15]. A general multi-agent framework is proposed in [16] to study the existence of and the convergence to CE in market interactions. Specifically, a strategic user is assumed to model the market price as a linear function of its desired demand. It is observed that it may be better or worse off than without modeling, depending on its initial belief. However, we note that using the linear model is purely heuristic in [16]. In contrast, we apply CE in the water-filling game, because it provides a practical solution concept to approach the performance bound of SE. We show that deploying the linear model to form conjectures can suitably explore the problem structure of the water-filling game, and therefore, lead to a substantial performance improvement.

The rest of the paper is organized as follows. Section II presents the non-cooperative game model, reviews the existing non-cooperative solutions, and introduces the concept of CE. The existence of this CE in the water-filling game is proved in Section III. Section IV develops practical algorithms to form beliefs and approach CE. Numerical simulations show that a foresighted user can achieve substantial performance improvement if it models its competitors in the water-filling game. Conclusions are drawn in Section V.

## II. System Model and Conjectural Equilibrium

In this section, we describe the mathematical model of the frequency-selective interference channel and formulate the non-cooperative multi-user water-filling game. We summarize the existing non-cooperative game theoretic solutions and introduce the conjecture equilibrium in the water-filling game context.

### A. System Description and Existing Solutions

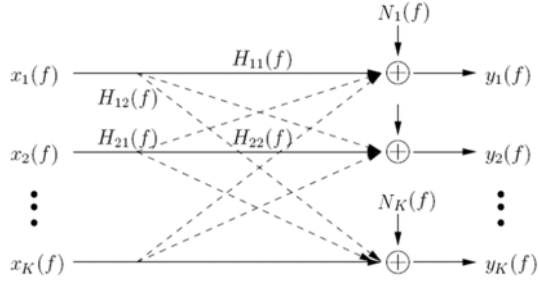

Fig. 1. Gaussian interference channel model.

Fig. 1 illustrates a frequency-selective Gaussian interference channel model. There are $K$ transmitters and $K$ receivers in the system. Each transmitter and receiver pair can be viewed as a player (or user). The transfer function of the channel from transmitter $i$ to receiver $j$ is denoted as $H_{ij}(f)$, where $0 \leq f \leq F_s$. The noise power spectral density (PSD) that receiver $k$ experiences is denoted as $N_k(f)$. Denote player $k$'s transmit PSD as $P_k(f)$. For user $k$, the transmit PSD is subject to its power constraint:

$$\int_0^{F_s} P_k(f) df \leq \mathbf{P_k^{max}}.$$ (1)

For a fixed $P_k(f)$, if treating interference as noise, user $k$ can achieve the following data rate:

$$R_k = \int_0^{F_s} \ln\left(1 + \frac{P_k(f)}{\sigma_k(f) + \sum_{j \neq k} P_j(f) \alpha_{jk}(f)}\right) df.$$ (2)

where $\sigma_k(f)$ and $\alpha_{jk}(f)$ are defined as $N_k(f) / |H_{kk}(f)|^2$ and $|H_{jk}(f)|^2 / |H_{kk}(f)|^2$.

To fully capture the performance tradeoff in the system, the concept of a rate region is defined as

$$\mathcal{R} = \left\{ (R_1, \cdots, R_K) : \exists (P_1(f), \cdots, P_K(f)) \ satisfying \ (1) \ and \ (2) \right\}.$$

The multi-user interaction in the interference channel can be modeled as a game. Let $\mathcal{G} = (\mathcal{K}, \mathcal{A}, U)$ denote a game with $\mathcal{K} = \{1, \cdots, K\}$ being the set of players, $\mathcal{A} = \times_{k \in \mathcal{K}} \mathcal{A}_k$ being the set of actions available to the users (in which $\mathcal{A}_k$ is the set of actions available to user $k$), and $U = \times_{k \in \mathcal{K}} U_k$ being the users' payoff functions (in which $U_k : \mathcal{A} \to \mathcal{R}$ is the user $k$'s payoff function) [13]. In the water-filling game, the players' payoffs are the respective achievable data rates and their strategies are to determine their transmit PSDs satisfying the constraint in (1).

As mentioned already in the introduction, existing research mainly focuses on two types of games, i.e. *cooperative* games and *non-cooperative* games. Specifically, cooperative approaches aim to maximize the weighted sum of data rates $\sum_{k=1}^{K} \omega_k R_k$. Because of the non-convexity in the rate expression as a function of power allocations, the computational complexity of optimal solutions (e.g., exhaustive search) in finding the rate region is prohibitively high. Existing works [9]-[11] aim to approach the Pareto boundary of this rate region and provide near-optimal performance. Moreover, it should be noted that cooperation among users is indispensable for the multi-user system to operate at the Pareto boundary. On the other hand, instead of solving the optimization problem globally, the IW algorithm models the users as myopic decision makers [1]. This means that they optimize their transmit PSD by water-filling and compete to increase their transmission rates with the sole objective of maximizing their own data rates in (2) regardless of the coupling among users. In other words, users are assumed to be myopic, i.e., they update actions shortsightedly without considering the long-term impacts of taking these actions. The outcome of this non-cooperative scenario is characterized by the concept of Nash equilibrium, which is defined to be any point $(a_1^*, \cdots, a_K^*)$ satisfying

$$U_k(a_k^*, a_{-k}^*) \geq U_k(a_k, a_{-k}^*) \text{ for all } a_k \in \mathcal{A}_k \text{ and } k \in \mathcal{K}, \tag{3}$$

where $a_{-k}^* = (a_1^*, \cdots, a_{k-1}^*, a_{k+1}^*, \cdots, a_K^*)$ [13]. The existence and the uniqueness of NE are proved under a wide range of realistic conditions and can be obtained by the IW algorithm [2]-[4].

The recent approach in [8] demonstrates that the myopic behavior can be further improved because this does not consider the coupling nature of players' actions and payoffs. If a selfish user gets the private information about its competitors and knows how they react, the best response strategy is to play the SE strategy. To define SE, we first define the action $a_k^*$ to be a best response (BR) to actions $a_{-k}$ if

$$U_k(a_k^*, a_{-k}) \geq U_k(a_k, a_{-k}), \forall\ a_k \in \mathcal{A}_k. \tag{4}$$

User $k$'s best response to $a_{-k}$ is denoted as $BR_k(a_{-k})$. Let $NE(a_k)$ be the Nash equilibrium strategy[1] of the remaining players if player $k$ chooses to play $a_k$, i.e.

$$NE(a_k) = a_{-k}, \text{ if } a_i = BR_i(a_{-i}) \text{ for any } i \neq k. \tag{5}$$

The strategy profile $(a_k^*, NE(a_k^*))$ is a Stackelberg equilibrium with user $k$ leading if and only if [14]

$$U_k(a_k^*, NE(a_k^*)) \geq U_k(a_k, NE(a_k)), \forall a_k \in \mathcal{A}_k. \tag{6}$$

Specifically, to find the SE in the water-filling game, we need to solve the following bi-level programming problem [8], where user 1 is assumed to be the foresighted user:

$$
\begin{aligned}
&\max_{P_1(f)} \int_0^{F_s} \ln\left(1 + \frac{P_1(f)}{\sigma_1(f) + \sum_{k=2}^{K} \alpha_{k1}(f) P_k(f)}\right) df \\
&s.t. \ \int_0^{F_s} P_1(f)\, df \leq \mathbf{P_1^{max}}, \\
&\quad\ \ P_1(f) \geq 0, \\
&\quad\ \ P_k(f) = \arg\max_{P_k'(f) \in \mathcal{A}_k} \int_0^{F_s} \ln\left(1 + \frac{P_k'(f)}{\sigma_k(f) + \sum_{i=1, i\neq k}^{K} \alpha_{ik}(f) P_i(f)}\right) df,\ k = 2, \cdots, K.
\end{aligned}
\tag{7}
$$

It should be pointed out that the foresighted user needs to know the private information $\sigma_k(f), \alpha_{ik}(f), \mathbf{P_k^{max}}$ of all its competitors in order to formulate the above optimization. The previous approach in [8] assumes that the foresighted user has the perfect knowledge of this private information. Importantly, it was shown in [8] that users' performance is substantially improved compared with that of IW algorithm if the foresighted user plays the SE strategy, even though the remaining users behave myopically. However, how such a foresighted user should accumulate this required information remains unsolved. In the remaining part of this paper, we will show that the foresighted user can obtain the information and improve its performance

---

[1]For the cases where the equilibrium solution $NE(a_k)$ is not unique for every $a_k$, the Stackelberg equilibrium needs to be reformulated [14]. Note that in this paper, $NE(a_k)$ is always unique in the water-filling game.

by forming conjectures over the behavior of its competitors through the repeated interaction.

Before introducing the conjectural equilibrium, we define the discretized version of the water-filling game. In practice, instead of optimizing over continuous frequency variables, the frequency band is often divided into a total number of $N$ small frequency bins [9]-[11], such that each frequency bin could be viewed as a flat fading channel and $\sigma_k(f), \alpha_{jk}(f)$ can be approximated as a constant within each small frequency bin. Denote $\sigma_k(f) = \sigma_k^n, \alpha_{jk}(f) = \alpha_{jk}^n,$ and $P_k(f) = P_k^n$ in $\frac{n-1}{N}F_s < f < \frac{n}{N}F_s$ for any $n \in \{1, \cdots, N\}, j, k \in \mathcal{K}$. As a result, (2) and (7) can be reformulated correspondingly.

### B. Conjectural Equilibrium

In game-theoretic analysis, conclusions about the reached equilibria are based on assumptions about what knowledge the players possess. For example, the standard NE solution assumes that every player knows its own payoff and believes that the other players' actions will not change. Therefore, it chooses to myopically maximize its own payoff [13]. Another example is that, to play the SE strategy, the foresighted user needs to know the structure of the resulting $NE(a_k)$ for any $a_k \in \mathcal{A}_k$ and believes that all the remaining players play the NE strategy. Summarizing, the players operating at equilibrium can be viewed as decision makers behaving optimally with respect to their *beliefs* about the policies of other players.

To rigorously define CE, we need to include two new elements $\mathcal{S}$ and $s$ and, based on this, reformulate the strategic game $\mathcal{G} = (\mathcal{K}, \mathcal{A}, U, \mathcal{S}, s)$ [16]. $\mathcal{S} = \times_{k \in \mathcal{K}} \mathcal{S}_k$ is the *state space*, where $\mathcal{S}_k$ is the part of the state relevant to the $k$ th user. Specifically, the state in the water-filling game is defined as the interference that users experience. The utility function $U = \times_{k \in \mathcal{K}} U_k$ is a map from users' state space to real numbers, $U_k : \mathcal{S}_k \times \mathcal{A}_k \to \mathcal{R}$. The *state determination function* $s = \times_{k \in \mathcal{K}} s_k$ maps joint action to state with each component $s_k : \mathcal{A} \to \mathcal{S}_k$. Each user cannot directly observe the actions chosen by the others, and each user has some belief about the state that would result from performing its available actions. The *belief function* $\tilde{s} = \times_{k \in \mathcal{K}} \tilde{s}_k$ is defined to be $\tilde{s}_k : \mathcal{A}_k \to \mathcal{S}_k$ such that $\tilde{s}_k(a_k)$ represents the state that the player $k$ believes that would result if it selects action $a_k$. Notice that the beliefs are not expressed in terms of other player's actions and preferences, and the multi-user coupling in these beliefs is captured directly by individual users forming conjectures of the effects of their own actions. In non-cooperative scenarios, each user chooses the action $a_k \in \mathcal{A}_k$ if it believes this action maximizes its utility.

**Definition 1 (*Conjectural Equilibrium*):** In the game $\mathcal{G}$ defined above, a configuration of belief functions $(\tilde{s}_1^*, \cdots, \tilde{s}_K^*)$ and a joint action $a^* = (a_1^*, \cdots, a_K^*)$ constitute a conjectural equilibrium, if for each $k \in \mathcal{K}$,

$$\tilde{s}_k^* \left( a_k^* \right) = s_k(a_1^*, \cdots, a_K^*) \text{ and } a_k^* = \arg\max_{a_k \in \mathcal{A}_k} U_k \left( \tilde{s}_k^* \left( a_k \right) \right). \tag{8}$$

From the definition, we can see that, at CE, all users' expectations based on their beliefs are realized and each user behaves optimally according to its expectation. In other words, users' beliefs are consistent with the outcome of the play and they behave optimally with respect to their beliefs. CE considers the users' beliefs rather than their perfect knowledge $NE(a_k)$ as in SE, which makes CE an appropriate solution concept when the perfect knowledge is not available. The key problem is how to configure the belief functions such that it leads to a CE having a satisfactory performance. Section III discusses this problem in water-filling games.

## III. Conjectural Equilibrium in Water-filling Games

In this section, we discuss how to configure a user's belief about its experienced interference as a linear function of its transmitted power, and show that such CE exists and it is a relaxation of both NE and SE. We begin by stating several fundamental assumptions used throughout the investigation hereafter.

*Assumption 1:* There is only one foresighted user modeling its competitors' reaction as a function of its allocated power, and all the remaining users are myopic users that deploy the IW algorithm. Without loss of generality, we assume that this foresighted user is user 1.

*Assumption 2:* Every user is able to perfectly measure its experienced equivalent noise PSD $\sigma_k^n$ and interference PSD $\sum_{i=1, i\neq k}^{K} \alpha_{ik}^n P_i^n$ in all frequency channels.

*Assumption 3*: Users $2, \cdots, K$ react to any small variation in their experienced interference by setting their power allocations according to the water-filling strategy.

*Assumption 4*: In the lower-level problem formed by user $2, \cdots, K$ in (7), there always exists a unique NE. For the sufficient conditions of the existence and uniqueness of NE, we refer readers to [2]-[4].

We formally define the concept of stationary interference.

*Definition 2 (**Stationary Interference**)*: The stationary interference that user 1 experiences in the $n$ th channel is the accumulated interference $I_1^n = \sum_{i=2}^{K} \alpha_{i1}^n P_i^n$ when best-response users $2, \cdots, K$ reach their NE in the lower-level problem in (7). Note that $I_1^n$ is in fact a function of user 1's power allocation $\boldsymbol{P_1} = \left[ P_1^1, \cdots, P_1^N \right]$ in the water-filling game and it can also be denoted as $I_1^n \left( \boldsymbol{P_1} \right)$.

## A. Linear Belief of Stationary Interference

As discussed before, both the state space and belief functions need to be defined in order to investigate the existence of CE. In the market models for pure exchange economy [16], the action set of each consumer is its desired demand based on the market price. The market price is also impacted by the other consumers' announced demand. Therefore, it is natural to define the state to be the market price in such scenarios. However, it should be pointed out that, for the problem considered in [16], modeling and updating the belief on the market price as a linear function of the excess demand is entirely heuristic. However, this is not the case in our setting, where forming linear conjectures is natural for the considered interference game.

In the water-filling game, we define state $\mathcal{S}_k$ to be the stationary interference caused to user $k$, because besides its own power allocation, its utility only depends on the interference that its competitors cause to it. Notice that the action available to user $k$ is to choose the transmitted power allocations subjected to its maximum power constraint. By the definition of belief function in Section II.B, we need to express the stationary interference as a function of the transmitted power. As we will see later, it is natural to deploy linear belief models due to the linearity of caused stationary interference in terms of the allocated power, and hence, forming such beliefs can lead to significant performance improvements because they capture the inherent characteristics of the actual interference coupling.

Define $\boldsymbol{P}_1^{n+}, \boldsymbol{P}_1^{n-}$ as $\boldsymbol{P}_1^{n+} = \left[ P_1^1, \cdots, P_1^n + \varepsilon, \cdots, P_1^N \right]$, $\boldsymbol{P}_1^{n-} = \left[ P_1^1, \cdots, P_1^n - \varepsilon, \cdots, P_1^N \right]$ for arbitrarily small positive variation in power $\varepsilon$. Given user 1's power allocation $\boldsymbol{P}_1$, $NE_n\left(\boldsymbol{P}_1\right) = \left[ P_2^n, \cdots, P_K^n \right]^T$ represents the power that user $2, \cdots, K$ allocate in the $n$ th channel at equilibrium. Vector $\boldsymbol{\alpha^n} = \left\{ \alpha_{ij}^n : i \neq j \right\}$ contains channel gains in the $n$ th frequency bin. Indicator function $\boldsymbol{y} = sign\left(\boldsymbol{x}\right)$ is a mapping of $\mathcal{R}^{K-1} \rightarrow \{0,1\}^{K-1}$, which is defined to be: $y_i = 1$, if $x_i > 0$, and $y_i = 0$, otherwise. Based on these notations, the following proposition motivates us to develop linear belief functions of stationary interference.

**Proposition 1** (*Linearity of Stationary Interference*)**:** If the number of frequency bins $N$ is sufficiently large, the first derivative of the stationary interference that user 1 experiences in the $n$ th channel with respect to its allocated power in the $m$ th channel satisfies

$$\frac{\partial I_1^n}{\partial P_1^n} = c\left(\boldsymbol{\alpha^n}, sign\left(NE_n\left(\boldsymbol{P}_1\right)\right)\right), \textit{if there does not exist } k \in \{2, \cdots, K\} \textit{ satisfying } P_k^n = 0 \textit{ and } \lambda_k^n = 0;$$

$$\left.\frac{\partial I_1^n}{\partial P_1^n}\right|_{P_1^n \rightarrow P_1^{n+}} = c\left(\boldsymbol{\alpha^n}, sign\left(NE_n\left(\boldsymbol{P}_1^{n+}\right)\right)\right), \left.\frac{\partial I_1^n}{\partial P_1^n}\right|_{P_1^n \rightarrow P_1^{n-}} = c\left(\boldsymbol{\alpha^n}, sign\left(NE_n\left(\boldsymbol{P}_1^{n-}\right)\right)\right), \textit{otherwise};$$

$$\frac{\partial I_1^n}{\partial P_1^m} = 0, \textit{if } m \neq n$$

in which $\lambda_k^n \left(k \in \{2, \cdots, K\}\right)$ is the Lagrange multiplier of $P_k^n \geq 0$ at the optimum of lower-level problem in (7). The function $sign\left(\cdot\right)$ is the indicator of which polyhedron the piece-wise affine water-filling function [3] lies in. $c\left(\boldsymbol{\alpha^n}, \boldsymbol{y}\right)$ represents a constant determined by $\boldsymbol{\alpha^n}$ and the non-zero elements of $\boldsymbol{y}$.

**Proof**: See Appendix A.

Proposition 1 indicates that, the first derivative with respect to a foresighted user's allocated power in a certain channel is sufficient to capture how the stationary interference varies locally in that channel. We observe from equality (11) that $I_1^n = \boldsymbol{h}^n \cdot NE_n\left(\boldsymbol{P}_1\right) = \boldsymbol{h}^n \left(\boldsymbol{I} + \boldsymbol{G}\right)^{-1} \boldsymbol{\nu} - \boldsymbol{h}^n \left(\boldsymbol{I} + \boldsymbol{G}\right)^{-1} \boldsymbol{g}^n P_1^n$. Therefore, user 1 can define its belief function using the linear form[2] $\tilde{I}_1^n = \beta^n - \gamma^n P_1^n$, in which $\gamma^n$ is the estimate of $-\dfrac{\partial I_1^n}{\partial P_1^n}$ and $\beta^n$ is a constant representing the composite effect of user $2, \cdots, K$ 's water-levels $\boldsymbol{\nu}$. This linear characterization of the stationary interference can greatly simplify the implicit functional expression $I_1^n\left(\boldsymbol{P_1}\right)$ given by the solution of lower-level problem (7), while maintaining an accurate model of $I_1^n\left(\boldsymbol{P_1}\right)$ around the feasible operating point $\boldsymbol{P_1}$.

---

[2] Note that as long as the channel realization is random, for a fixed $N$, the probability that the left-sided and right-sided derivatives in proposition 1 are not equal is zero. We will assume that the first derivative exists hereafter. If it does not exist, similar results can be derived by treating the left-sided and right-sided first derivatives separately.

### B. Existence of Conjectural Equilibrium

Under the same known sufficient conditions discussed in [3][4][8] for guaranteeing the existence of NE and SE, the existence of CE can be proved by showing that the first two types of equilibrium are special cases of CE. To this end, Table I compares the optimality conditions of the three types of equilibria in the water-filling game.

| | User 1 | User $2,\cdots,K$ |
|---|---|---|
| Nash Equilibrium | $\{P_k^n\} = \arg\max\limits_{\{P_k^n\}\in\mathcal{A}_k} \sum\limits_{n=1}^{N}\log_2\left(1+\dfrac{P_k'^n}{\sigma_k^n+I_k^n}\right)$ | |
| Stackelberg Equilibrium | $\{P_1^n\} = \arg\max\limits_{\{P_1^n\}\in\mathcal{A}_1} \sum\limits_{n=1}^{N}\log_2\left(1+\dfrac{P_1'^n}{\sigma_1^n+I_1^n\left(\boldsymbol{P'}\right)}\right)$ | $\{P_k^n\} = \arg\max\limits_{\{P_k^n\}\in\mathcal{A}_k} \sum\limits_{n=1}^{N}\log_2\left(1+\dfrac{P_k'^n}{\sigma_k^n+I_k^n}\right)$ |
| Conjectural Equilibrium | $\{P_k^n\} = \arg\max\limits_{\{P_k^n\}\in\mathcal{A}_k} \sum\limits_{n=1}^{N}\log_2\left(1+\dfrac{P_k'^n}{\sigma_k^n+\tilde{I}_k^n}\right)$ | |
| | $\tilde{I}_1^n = \beta^n - \gamma^n P_1^n, I_1^n = \sum\limits_{i=2}^{K}\alpha_{i1}^n P_i^n, \tilde{I}_1^n = I_1^n$ | $\tilde{I}_k^n = I_k^n = \sum\limits_{i=1,i\neq k}^{K}\alpha_{ik}^n P_i^n$ |

Table I. Comparison among NE, SE, and CE in water-filling games.

As shown in Table I, the information requirement for playing various equilibria differs. At NE, each user includes its stationary interference $I_k^n$ as a constant in the optimization's, and its action is the best response to $I_k^n$. To play SE, the foresighted user needs to know the functional expression of the stationary interference $I_1^n\left(\boldsymbol{P'}\right)$ such that the bi-level program can be formed. Specifically, the required information includes both the system-wide channel state information $\boldsymbol{\alpha^n}$, the noise PSD $\sigma_k^n$, and the individual power constraint $\mathbf{P_k^{max}}$ for $\forall n \in \{1,\cdots,N\}, k \in \mathcal{K}$. In contrast, in the case of CE, the above information for playing SE is no longer required and the foresighted user behaves optimally with respect to its beliefs $\tilde{I}_1^n$ on how the stationary interference changes as a function of $P_1^n$.

***Proposition 2** (**NE & SE as CE**):* Both Nash equilibrium and Stackelberg equilibrium are special cases of conjectural equilibrium.

***Proof***: In order to show that both NE and SE are special cases of CE, we only need to verify that at NE and SE, user 1's action is optimal with respect to its belief and its belief agrees with its state. First, clearly, NE is a trivial CE with the parameters $\beta^n = \sum_{i=2}^{K} \alpha_{i1}^n P_i^n, \gamma^n = 0$ in user 1's belief functions. Next, denote $P_{\text{SE}} = \left[ P_{SE}^1, \cdots, P_{SE}^N \right]$ the optimal solution of the discretized version of problem (7). To prove that SE is a CE, we need to find the corresponding $\beta^n$ and $\gamma^n$ and show that SE also solves problem (13). Consider the belief function in Table I with the parameters $\beta^n = \left( I_1^n - P_1^n \cdot \dfrac{\partial I_1^n}{\partial P_1^n} \right) \Bigg|_{P_1 = P_{\text{SE}}}$ and $\gamma^n = -\dfrac{\partial I_1^n}{\partial P_1^n} \Bigg|_{P_1 = P_{\text{SE}}}$. As discussed before, such parameters preserve all the local information of the objective of problem (7) around $P_{\text{SE}}$ into problem (13). KKT conditions hold at $P_{\text{SE}}$ since it solves problem (7). A sufficient condition which ensures SE to be a CE is that problem (13) belongs to convex optimization, because KKT conditions are necessary and sufficient for convex programming to attain its optimum. Appendix B provides a sufficient condition **SC 1** under which problem (13) is convex, thereby proving that SE is a special CE given these conditions. ∎

Proposition 2 indicates that the two isolating points, NE and SE, are both CE, if parameters $\beta = \left\{ \beta^n \right\}$, $\gamma = \left\{ \gamma^n \right\}$ are properly chosen. Therefore, CE can be viewed as an operational approach to attain SE if the system-wide information required for solving SE is not available. It is because only the local information including stationary interference $I_1^n$ and its first derivative $\dfrac{\partial I_1^n}{\partial P_1^n}$ is required to formulate problem (13), and this information can be obtained using measurements performed at the receiver.

In addition, we are interested in the existence of other CEs besides these two points. Denote the parameters of any CE, e.g. NE or SE, as $\beta_* = \left\{ \beta_*^n \right\}$, $\gamma_* = \left\{ \gamma_*^n \right\}$, and the optimal solution of problem (13) given parameters $\beta, \gamma$ as $P_1 \left( \beta, \gamma \right)$. Let $F : \mathcal{R}^N \times \mathcal{R}^N \to \mathcal{R}^N$ be a mapping defined as $F \left( \beta, \gamma \right) =$

$\{F^n(\boldsymbol{\beta}, \boldsymbol{\gamma})\}$ in which

$$F^n(\boldsymbol{\beta}, \boldsymbol{\gamma}) = \boldsymbol{h}^n \cdot NE_n(\boldsymbol{P_1}(\boldsymbol{\beta}, \boldsymbol{\gamma})) - \beta^n - \gamma^n P_1^n(\boldsymbol{\beta}, \boldsymbol{\gamma}). \tag{9}$$

The following proposition gives a sufficient condition which ensures that infinite CEs exist.

***Proposition 3 (Infinite Set of CE)*:** Let $\mathcal{G}$ be a water-filling game that satisfies condition $\mathsf{SC\ 1}$. Suppose that all the users form conjectures according to Table I. If there exist open neighborhoods $A \subset \mathcal{R}^N$ and $B \subset \mathcal{R}^N$ of $\boldsymbol{\beta}_*$ and $\boldsymbol{\gamma}_*$ respectively, such that $F(\bullet, \boldsymbol{\gamma}) : A \to \mathcal{R}^N$ is locally one-to-one for any $\boldsymbol{\gamma} \in B$, then $\mathcal{G}$ admits an infinite set of conjectural equilibria.

***Proof***: See Appendix C.

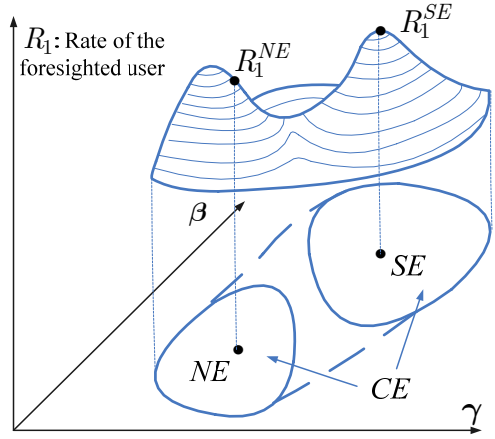

Fig. 2. Structure of conjectural equilibria in water-filling games.

In summary, proposition 1, 2, and 3 characterize the existence and structure of conjectural equilibrium in water-filling games. As shown in Fig. 2, NE and SE can be both special cases of CE. Open sets of CE that contain NE and SE may exist in the $\boldsymbol{\beta}$ - $\boldsymbol{\gamma}$ plane and these sets may be connected. SE attains the maximal data rate that a foresighted user can achieve. If the foresighted user properly sets up its parameters $\beta, \gamma$, the solution of CE in problem (13) coincides with the solution of SE in problem (7). More importantly, as opposed to the SE in which the knowledge of the system-wide private information is required, CE assumes that the foresighted user knows only its stationary interference and the first derivatives with respect to the allocated power, which greatly reduces the complexity of information acquisition. Therefore, in order to approach the performance upper bound given by SE, this paper adopts the approach of CE. Section IV will develop practical conjecture forming and updating algorithms to select out of the infinite CEs a desirable power allocation scheme that provides comparable achievable rates with SE.

## IV. Achieving Desirable Conjectural Equilibria

Since proposition 3 shows that infinite CEs may exist and SE is the most desirable CE for a foresighted user, it should wisely choose the parameters $\beta^n, \gamma^n$ of belief functions in order to attain SE as a CE. Moreover, the declarative conclusions drawn in Section III provide no hint on the achievability of the CE. However, in practice, it is more important to construct algorithmic mechanisms to attain the desirable CE.

To arrive at CE, a multi-agent learning approach is proposed for the repeated game setting [16]. Let $\tilde{s}_{k,t}$ and $a_{k,t}$ denote user $k$'s belief and action at time $t$. In the framework, at time $t$, the users update their beliefs $\tilde{s}_{k,t}$ and select their actions $a_{k,t}$ based on their past observations. A learning algorithm converges if $\lim_{t\to\infty}\left(\tilde{s}_{1,t},\cdots,\tilde{s}_{K,t}\right)$ and $\lim_{t\to\infty}\left(a_{1,t},\cdots,a_{K,t}\right)$ is a CE. If we define learning as the players' dynamic process of building beliefs by forming conjectures about the effects of their actions, CE captures the achieved outcome when consistency of conjectures within and across players emerges.

Similarly, this section proposes that users can update their beliefs in the repeated interaction setting and numerically examines their performance. Before going into the technical details, it should be pointed out that the pursuit of the practical solution's convergence to CE is not the principal goal of our investigation. Instead, computing power allocation strategies that require only local information and achieve comparable rates with SE (which requires global information) is the ultimate objective rather than the convergence. In other words, any power allocation strategy that lies outside the open CE set in Fig. 2 is favorable if it can improve the performance compared with NE.

### A. Conjecture-based Rate Maximization

Table II summarizes the dynamic updates of all users' states, belief functions, and optimal actions in the water-filling game. Specifically, at iteration $t$, users' states $I_{k,t}$ are determined by their opponents' power allocation. User 1 updates the parameters $\beta_t^n, \gamma_t^n$ in its belief functions based on its state $I_{1,t}^n$ and allocated power $P_{1,t}^n$, and it also updates its power allocation $\boldsymbol{P}_{1,t+1}$ based on current operating points $\boldsymbol{P}_{1,t}$ and its belief $\tilde{\boldsymbol{I}}_{1,t}$. At the same time, myopic users $2, \cdots, K$ set their belief equal to their experienced interference and update their power allocation based on the water-filling strategy. Note that Table II implicitly assumes that user 1 will update after user $2, \cdots, K$ 's IW algorithms converge such that user $2, \cdots, K$ 's power allocations $\boldsymbol{P}_{k,t}$ at time $t$ can be regarded as an equilibrium state. The outcome of this dynamic play is a CE if $\lim_{t \to \infty} \boldsymbol{P}_{k,t}$ exists and $\lim_{t \to \infty} \boldsymbol{I}_{k,t} = \lim_{t \to \infty} \tilde{\boldsymbol{I}}_{k,t}$. As discussed in the proof of proposition 2, it is equivalent to check the convergence of user 1's updates. We can see from Table II that user 1 needs to complete two updates at each iteration. The entire procedure in Table II that enables the foresighted user to build beliefs and improve its performance is named "*Conjecture-based Rate Maximization*". Appropriate rules for updating beliefs are discussed as follows.

| | User 1 | User $2, \cdots, K$ |
|---|---|---|
| State $I_{k,t}$ | $I_{k,t}^n = \sum_{i=1, i \neq k}^{K} \alpha_{i1}^n P_{i,t}^n$ | |
| Belief function $\tilde{s}_k : \mathcal{A}_k \to \mathcal{S}_k$ | $\beta_t^n, \gamma_t^n \leftarrow \textbf{Update}_1\left(I_{1,t}^n, P_{1,t}^n\right)$ $\tilde{I}_{1,t}^n = \beta_t^n - \gamma_t^n P_{1,t}^n$ | $\tilde{I}_{k,t}^n = I_{k,t}^n = \sum_{i=1, i \neq k}^{K} \alpha_{ik}^n P_{i,t}^n$ |
| Action $a_{1,t}, \cdots, a_{K,t}$ | $\boldsymbol{P}_{1,t+1} \leftarrow \textbf{Update}_2\left(\boldsymbol{P}_{1,t}, \tilde{\boldsymbol{I}}_{1,t}\right)$ | $\boldsymbol{P}_{k,t} = \arg \max_{\boldsymbol{P}_k^t \in \mathcal{A}_k} \sum_{n=1}^{N} \log_2\left(1 + \dfrac{P_k^{tn}}{\sigma_k^n + \tilde{I}_{k,t}^n}\right)$ |

<p align="center">Table II. Dynamic updates of the play.</p>

**Update$_1$:** $\beta_t^n, \gamma_t^n$

Note that, we have $I_1^n = \boldsymbol{h}^n \left(\boldsymbol{I} + \boldsymbol{G}\right)^{-1} \boldsymbol{\nu} - \boldsymbol{h}^n \left(\boldsymbol{I} + \boldsymbol{G}\right)^{-1} \boldsymbol{g}^n P_1^n$ from proposition 1, user 1's belief function takes the form of $\tilde{I}_1^n = \beta^n - \gamma^n P_1^n$, and it satisfies $I_1^n = \tilde{I}_1^n$ at CE for any $n \in \{1, \cdots, N\}$. As discussed in the previous section, by setting the parameters $\beta^n = I_1^n - P_1^n \cdot \dfrac{\partial I_1^n}{\partial P_1^n}$ and $\gamma^n = -\dfrac{\partial I_1^n}{\partial P_1^n}$, we can preserve all the local information of the original SE problem (7) around current feasible operating point $\boldsymbol{P}_{1,t}$. Therefore, we can update $\beta_t^n$ and $\gamma_t^n$ using $\beta_t^n = \left(I_1^n - P_1^n \cdot \dfrac{\partial I_1^n}{\partial P_1^n}\right)\Bigg|_{P_1 = P_{1,t}}$ and $\gamma_t^n = -\dfrac{\partial I_1^n}{\partial P_1^n}\Bigg|_{P_1 = P_{1,t}}$. By assumption 3, user 1

can approximate the parameters using $\dfrac{\partial I_1^n}{\partial P_1^n} \approx \dfrac{I_1^n\left(\left\{P_1^n + \varepsilon\right\} \cup \boldsymbol{P_1^{-n}}\right) - I_1^n\left(\left\{P_1^n - \varepsilon\right\} \cup \boldsymbol{P_1^{-n}}\right)}{2\varepsilon}$ for small $\varepsilon$ in which $\boldsymbol{P_1^{-n}} = \left\{P_1^1, \cdots, P_1^{n-1}, P_1^{n+1}, \cdots, P_1^N\right\}$.

After $\mathbf{Update_1}$ in each iteration, user 1 needs to solve problem (13). If proposition 2's assumption is not satisfied, problem (13) belongs to the class of non-convex optimization, which is generally hard to solve and standard optimization algorithms can only be used to determine local maxima [17]. However, in this application, we are able to show that, as long as the number of frequency bins $N$ is sufficiently large, problem (13) satisfies the time-sharing condition [10], and its global optimum can be efficiently computed.

***Definition 3 (Time-sharing Condition [10]):*** Consider an optimization problem with the general form:

$$\max \sum_{n=1}^{N} o_n\left(\mathbf{x}_n\right)$$
$$s.t. \sum_{n=1}^{N} \mathbf{c}_n\left(\mathbf{x}_n\right) \leq \mathbf{P}, \tag{10}$$

where $o_n\left(\mathbf{x}_n\right)$ are objective functions that are not necessarily concave, $\mathbf{c}_n\left(\mathbf{x}_n\right)$ are constraint functions that are not necessarily convex. Power constraints are denoted by $\mathbf{P}$. Let $\mathbf{x}_n^*$ and $\mathbf{y}_n^*$ be optimal solutions to the optimization problem (10) with $\mathbf{P} = \mathbf{P_x}$ and $\mathbf{P} = \mathbf{P_y}$, respectively. An optimization problem of the form (10) is said to satisfy the time-sharing condition if for any $0 \leq v \leq 1$, there always exists a feasible solution $\mathbf{z}_n$, such that $\sum_{n=1}^{N} \mathbf{c}_n\left(\mathbf{z}_n\right) \leq v\mathbf{P_x} + (1-v)\mathbf{P_y}$, and $\sum_{n=1}^{N} o_n\left(\mathbf{z}_n\right) \leq v\sum_{n=1}^{N} o_n\left(\mathbf{x}_n^*\right) + (1-v)\sum_{n=1}^{N} o_n\left(\mathbf{y}_n^*\right)$.

The following proposition indicates that problem (13) satisfies the time-sharing condition.

***Proposition 4 (Satisfaction of Time-sharing Condition):*** As the total number of sub-carriers $N$ goes to infinity, problem (13) satisfies the time-sharing condition.

***Proof:*** Specifically, for the optimization problem (13), $o_n\left(\mathbf{x}_n\right) = f_{\sigma_1^n, \beta^n, \gamma^n}\left(P_1^n\right)$, $\mathbf{c}_n\left(\mathbf{x}_n\right) = P_1^n$, $\mathbf{P} = \mathbf{P_1^{max}}$. First, consider the continuous case in which $\sigma_k\left(f\right)$ and $\alpha_{jk}\left(f\right)$ are both continuous functions of $f$. Set the initial value $\beta_0(f), \gamma_0(f)$ to be piece-wise continuous. By rule **Update$_1$**, $\beta_t(f)$ and $\gamma_t(f)$ are piece-wise continuous, because Appendix C proves that $P_1(\beta, \gamma)$ and $F(\beta, \gamma)$ are continuous in $(\beta, \gamma)$. We have the piece-wise continuity of $\beta(f), \gamma(f)$. The rest of the proof is similar to the proof of theorem 2 in [10]. ■

## **Update$_2$:** $P_{1,t+1}$

It is shown in [10] that, if the optimization problem satisfies the time-sharing property, then it has a zero duality gap, which leads to efficient numerical algorithms that solve the non-convex problem in the dual domain. Consider the dual objective function $d(\boldsymbol{\eta}) = \sum_{n=1}^{N} \left\{ \max_{P_1^n} f_{\sigma_1^n, \beta^n, \gamma^n}(P_1^n) - \boldsymbol{\eta} P_1^n \right\} + \boldsymbol{\eta} \mathbf{P_1^{max}}$. Since $d(\boldsymbol{\eta})$ is convex, a bisection or gradient-type search over the Lagrangian dual variable $\boldsymbol{\eta}$ is guaranteed to converge to the global optimum. Specifically, Algorithm 1 summarizes such a dual method that solves non-convex problem (13) using bisection update. As long as the time-sharing condition is satisfied, Algorithm 1 converges to the global optimum. Hence, we can always solve problem (13) regardless of its convexity.

---

**Algorithm 1 :** A dual method that solves problem (13) using bisection update

---

**input:** $\{\sigma_1^n\}, \{\beta_t^n\}, \{\gamma_t^n\}, \mathbf{P_1^{max}}$

**initialization :** $\boldsymbol{\eta}_{\min}, \boldsymbol{\eta}_{\max}, \boldsymbol{\eta}_0 = (\boldsymbol{\eta}_{\min} + \boldsymbol{\eta}_{\max})/2, i = 0$

**repeat**

set $\boldsymbol{P_1} = \left[ P_1^1 \cdots P_1^N \right]$ where $P_1^n = \arg \max_{P_1'^n \in \boldsymbol{dom} f_{\sigma_1^n, \beta^n, \gamma^n}} \log_2 \left( 1 + \dfrac{P_1'^n}{\sigma_1^n + \beta_t^n - \gamma_t^n P_1'^n} \right) - \boldsymbol{\eta}_i P_1'^n$.

if $\sum_n P_1^n < \mathbf{P_1^{max}}$, $\boldsymbol{\eta}_{\max} = \boldsymbol{\eta}_i$; else $\boldsymbol{\eta}_{\min} = \boldsymbol{\eta}_i$.

$\boldsymbol{\eta}_{i+1} \leftarrow (\boldsymbol{\eta}_{\min} + \boldsymbol{\eta}_{\max})/2, i = i + 1$.

**until** $\boldsymbol{\eta}_i$ converges

---

Table III. A dual algorithm that solves problem (13).

Table IV summarizes the procedure of algorithm "Conjecture-based Rate Maximization" (CRM). Next, we make several remarks about this algorithm. First, since we want to achieve better performance than NE, the initial operating point $P_{1,0}$ is set to be the power allocation strategy $P_1^{NE}$ that user 1 will choose if it adopts the IW algorithm. Second, in $\mathbf{Update_2}$, the global optimum $\boldsymbol{P}_1^c$ is not directly used to update $P_{1,t+1}^n$. As shown in Fig. 3, this is because problem (13) is only a local approximation at $\boldsymbol{P}_{1,t}$ of the original SE problem (7) that we want to solve. Using $\boldsymbol{P}_1^c$ to update $\boldsymbol{P}_{1,t+1}$ may decrease the actual achievable rate $R_1$, if a mismatch between problem (7) and (13) exists for the solution $\boldsymbol{P}_1^c$. Therefore, $\mathbf{Update_2}$ adopts line search to guarantee that the achievable rate is increased at each iteration. Third, as opposed to the two-user algorithm proposed in [8], CRM is designed for the general multi-user scenario regardless of the number of users. Last, CRM is not guaranteed to converge. It may stop with $v = 0$ and $\boldsymbol{P}_{1,t} \neq \boldsymbol{P}_1^c$ in $\mathbf{Update_2}$, i.e. the maximizer of $R_1$ in the interval between $\boldsymbol{P}_{1,t}$ and $\boldsymbol{P}_1^c$ in Fig. 3 is $\boldsymbol{P}_{1,t}$. However, it can be easily verified that, if $\boldsymbol{P}_{1,t} = \boldsymbol{P}_1^c$, CRM converges and the resulting outcome is a CE.

---

**Conjecture-based Rate Maximization**

---

**initialization :** $t = 0, P_{1,0} = P_1^{NE}$

**repeat**

   I.  $\beta_t^n, \gamma_t^n \leftarrow \mathbf{Update_1}\left(I_{1,t}^n, P_{1,t}^n\right)$.

   II. $P_{1,t+1} \leftarrow \mathbf{Update_2}\left(\boldsymbol{P}_{1,t}, \tilde{\boldsymbol{I}}_{1,t}\right)$, which includes:

      1) Consider problem $\max \sum_{n \in \mathcal{F}_1'} f_{\sigma_1^n, \beta^n, \gamma^n}\left(P_1^n\right), s.t. \ P_1^n \in \boldsymbol{dom} \ f_{\sigma_1^n, \beta^n, \gamma^n} \ and \ \sum_{n=1}^{N} P_1^n \le \mathbf{P_1^{max}}$.

      2) Use Algorithm 1 to calculate the global optimum $\boldsymbol{P}_1^c$ of the above problem.

      3) Search in the interval of $v\boldsymbol{P}_{1,t} + (1-v)\boldsymbol{P}_1^c \ (0 \le v \le 1)$ and find in the interval the power

         allocation $\boldsymbol{P}_1^s$ that maximizes user 1's actual achievable rate $R_1$.

      4) $\boldsymbol{P}_{1,t+1} \leftarrow \boldsymbol{P}_1^s, t = t + 1$.

**until** no improvement can be made.

---

Table IV.  Conjecture-based rate maximization.

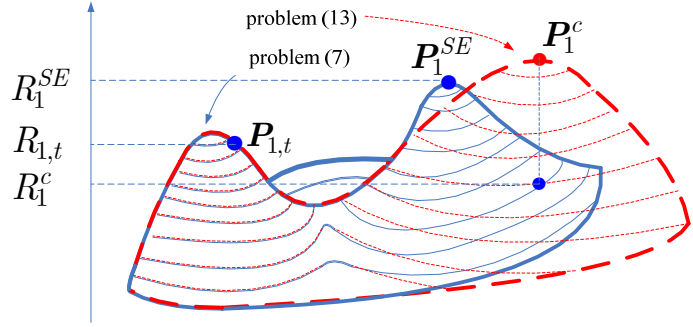

Fig. 3. Mismatch between problem (7) and (13).

### B. Illustrative Numerical Examples

This sub-section compares the performance of CRM with the IW algorithm and a two-user suboptimal algorithm (TSA) that searches SE assuming perfect knowledge of its opponent's private information [8]. We simulate a system with 200 sub-carriers over the 10-MHz band. We consider frequency-selective channels using a four-ray Rayleigh model with the exponential power profile and 100 ns root mean square delay spread. The power of each ray is decreasing exponentially according to its delay.

We first simulate the two-user scenario with $P_1 = P_2 = 200$ and $\sigma_1(f) = \sigma_2(f) = 0.01$. The total power of all rays of $H_{11}(f)$ and $H_{22}(f)$ is normalized as one, and that of $H_{12}(f)$ and $H_{21}(f)$ is normalized as $0.5$. Fig. 4 shows an example of user 1's power allocations when deploying different algorithms under the same conditions. In IW algorithm, user 1 water-fills the whole frequency band by regarding its competitor's interference as background noise. In contrast, user 1 will not water-fill if choosing CRM and TSA. It avoids the myopic behavior and improves its performance by explicitly considering the stationary interference caused by its opponent.

To evaluate the performance, we tested $10^5$ sets of frequency-selective fading channels where the Nash equilibrium exists. Denote user $i$'s achievable rate using CRM, IW and TSA as $R_i$, $R_i^{NE}$, and $R_i'^{SE}$ respectively. Fig. 5 shows the simulated cumulative probability of the ratio of $R_i$ over $R_i^{NE}$ and $R_i'^{SE}$. The curve indicates that there is a probability of 59% that CRM returns the same power allocation strategy as IW. On the other hand, the average improvement for user 1 of CRM over IW is 16.8%, which achieves almost the same performance as TSA. As shown in Fig. 5, $R_1 / R_1'^{SE}$ is distributed symmetrically with respect to $R_1 = R_1'^{SE}$. CRM provides for user 2 an average improvement of 20.7% over IW, which is smaller than TSA. Similarly as in [8], only in very few cases, CRM will result in a rate $R_2'$ smaller than $R_2^{NE}$ in the IW algorithm.

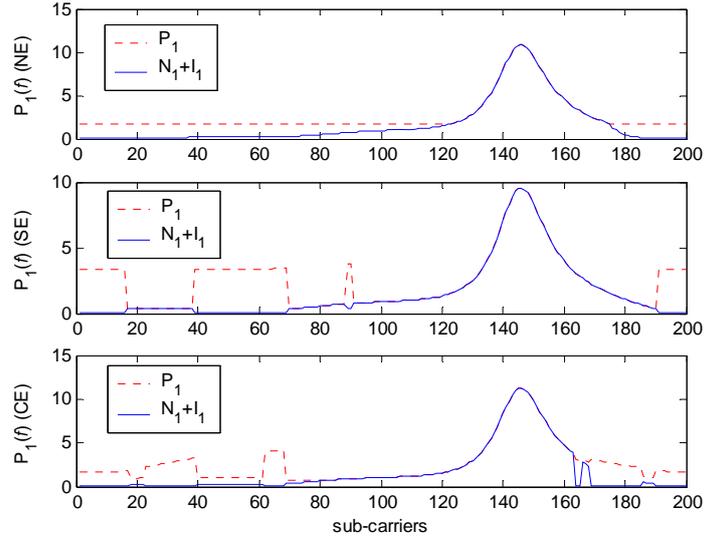

Fig. 4. User 1's power allocation using different algorithms.

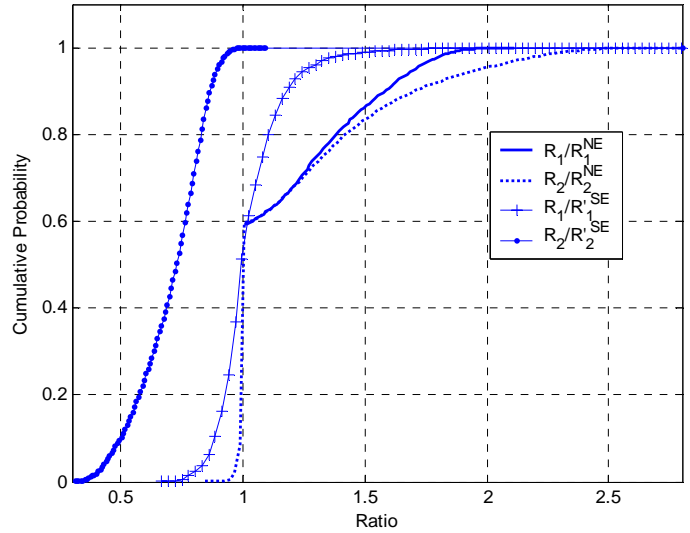

Fig. 5. Cdfs of $R_i / R_i^{NE}$ and $R_i / R_i'^{SE}$ $(i = 1, 2)$.

The iteration time required by CRM is summarized in Table V. As mentioned above, CRM stops after just one iteration with a probability of 59% due to the problem mismatch shown in Fig. 3. In most scenarios, CRM terminates within 4 iterations and the average number of required iteration is only 1.84. To further improve the performance of CRM, we can modify the original CRM to handle the problem mismatch between (7) and (13). Notice that problem (13) is only a local approximation of problem (7) at $\boldsymbol{P}_{1,t}$. Additional constraints can be added in Algorithm 1, such that the optimum of problem (13) is searched only in a certain region around $\boldsymbol{P}_{1,t}$ rather than the whole domain of $f_{\sigma_t^n, \beta_t^n, \gamma_t^n}$. For example, $\left|P_1'^n - P_{1,t}^n\right|$ can be restricted within a certain threshold when performing Algorithm 1 for any $n \in \{1, \cdots, N\}$. We simulated the two-user scenarios with additional restriction of $\left|P_1'^n - P_{1,t}^n\right| \le 1$. Fig. 6 shows the simulated cumulative probability of $R_i / R_i^{NE}$ for this modified CRM. As opposed to CRM, the probability that the modified CRM returns the same power allocation strategy as IW is reduced to 36% and the average performance improvement is also increased for both users. Specifically, the average performance improvement for user 1 is 24.4% and that of user 2 is 33.6%. However, Table V shows that the improvement is achieved at the cost of more iterations.

| | Probability of required iterations | | | | |
|---|---|---|---|---|---|
| | $t = 1$ | $t = 2$ | $t = 3$ | $t = 4$ | $t \ge 5$ |
| CRM | 0.59 | 0.07 | 0.26 | 0.07 | 0.01 |
| Modified CRM | 0.34 | 0.05 | 0.40 | 0.19 | 0.02 |

Table V.  Iterations required by different CRM algorithms.

We also tested performance of modified CRM in multi-user cases where TSA cannot be applied. We simulated the three-user scenarios with $\boldsymbol{P_k} = 200$ and $\sigma_k(f) = 0.01$. The total power of all rays of $H_{kk}(f)$ is normalized as one, and that of $H_{ij}(f)$ $(i \ne j)$ is normalized as $0.33$. Fig. 7 shows the simulated cumulative probability of $R_i / R_i^{NE}$. The average improvement for user 1 of modified CRM over IW is 26.3%, and that of the rest users is 9.7%. We can see that, it benefits most of the participants in the water-filling game if a foresighted user forms accurate conjectures and plays the conjecture equilibrium strategy.

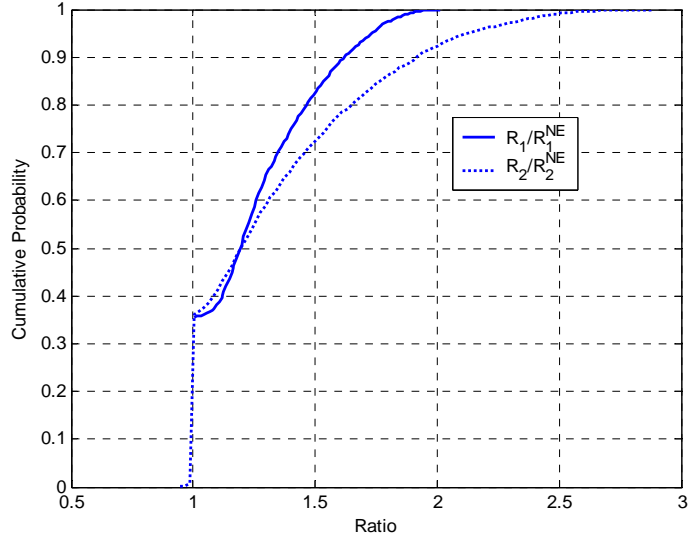

Fig. 6. Cdfs of $R_i \, / \, R_i^{NE}$ $(i = 1, 2)$ for modified CRM.

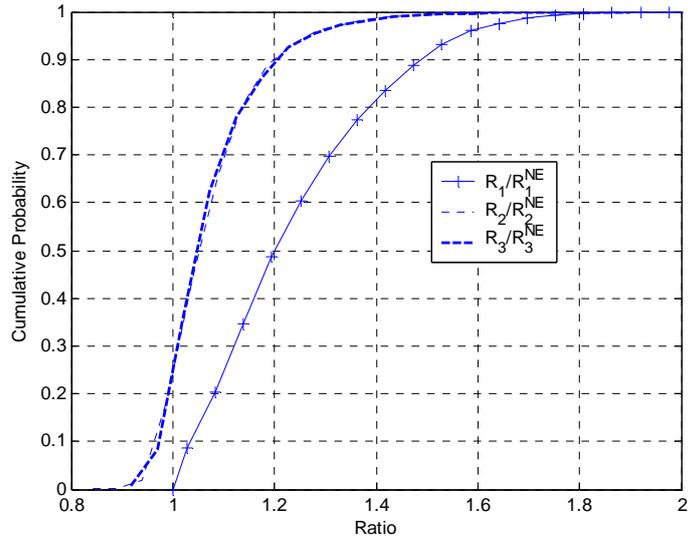

Fig. 7. Cdfs of $R_i \, / \, R_i^{NE}$ $(i = 1, 2, 3)$ for modified CRM.

## V. CONCLUSION

This paper introduces the concept of conjectural equilibrium in non-cooperative water-filling games and discusses how a foresighted user can model its experienced interference as a function of its own power allocation in order to improve its own data rate. The existence of conjectural equilibrium is proved and both game theoretic solutions, including Nash equilibrium and Stackelberg equilibrium, are shown to be special cases of this conjectural equilibrium. Practical algorithms based on conjectural equilibrium are developed to determine desirable power allocation strategies. Numerical results verify that a foresighted user forming proper conjectures can improves both its own achievable rate as well as the rates of other participants, even if it has no a priori knowledge of its competitors' private information.

## APPENDIX A

*Proof of Proposition 1*

By the definition of $I_1^n$, we have $\dfrac{\partial I_1^n}{\partial P_1^m} = \sum\limits_{i=2}^{K} \alpha_{i1}^n \dfrac{\partial P_i^n}{\partial P_1^m}$. We differentiate two different cases:

*1)* If there does not exist any $k \in \{2,\cdots,K\}$ satisfying $P_k^n = 0$ and $\lambda_k^n = 0$, i.e. there is none-zero gap between the interference that users $2,\cdots,K$ experiences and their water-levels, it is straightforward to see that $sign\left(NE_n\left(\boldsymbol{P_1}\right)\right) = sign\left(NE_n\left(\boldsymbol{P_1^{n+}}\right)\right) = sign\left(NE_n\left(\boldsymbol{P_1^{n-}}\right)\right)$.

Without loss of generality, we temporarily assume that $P_k^n > 0$ for $k \in \{2,\cdots,K\}$. When users $2,\cdots,K$ reach the equilibrium, we have from the optimality conditions of water-filling solution:

$$\left(\boldsymbol{I} + \boldsymbol{G}\right) \cdot NE_n\left(\boldsymbol{P_1}\right) + \boldsymbol{g}^n P_1^n = \boldsymbol{\nu}\,,$$

in which $\boldsymbol{G} = \begin{bmatrix} 0 & \alpha_{32}^n & \alpha_{42}^n & \cdots & \alpha_{K2}^n \\ \alpha_{23}^n & 0 & \alpha_{43}^n & \cdots & \alpha_{K3}^n \\ \alpha_{24}^n & \alpha_{34}^n & \ddots & \cdots & \vdots \\ \vdots & \vdots & \vdots & 0 & \alpha_{K,K-1}^n \\ \alpha_{2K}^n & \alpha_{3K}^n & \cdots & \alpha_{K-1,K}^n & 0 \end{bmatrix}, \boldsymbol{g}^n = \begin{bmatrix} \alpha_{12}^n \\ \alpha_{13}^n \\ \vdots \\ \alpha_{1K}^n \end{bmatrix}, \boldsymbol{\nu} = \begin{bmatrix} \nu_2 \\ \nu_3 \\ \vdots \\ \nu_K \end{bmatrix}, \nu_i\,(i=2,\cdots,K)$ are the water-levels of all the water-filling users. Note that the sufficient conditions of existence and uniqueness of NE generally require $\|\boldsymbol{G}\| < 1$ [3], which leads to the fact that $\boldsymbol{I} + \boldsymbol{G}$ is invertible. Therefore, we have

$$NE_n\left(\boldsymbol{P_1}\right) = \left(\boldsymbol{I} + \boldsymbol{G}\right)^{-1} \boldsymbol{\nu} - \left(\boldsymbol{I} + \boldsymbol{G}\right)^{-1} \boldsymbol{g}^n P_1^n\,. \tag{11}$$

We also have $\lim_{N \to \infty} \frac{\partial \nu_i}{\partial P_1^n} = 0$, because if the width of each frequency bin $F_s / N$ is sufficiently small, the fluctuation of the water-level is negligible. In other words, if $N$ is sufficiently large, we have $\frac{\partial \nu_i}{\partial P_1^n} \approx 0$. As a result, we have

$$\frac{\partial I_1^n}{\partial P_1^m} = \frac{\partial \boldsymbol{h}^n \cdot NE_n\left(\boldsymbol{P_1}\right)}{\partial P_1^m} = \begin{cases} -\boldsymbol{h}^n \left(\boldsymbol{I} + \boldsymbol{G}\right)^{-1} \boldsymbol{g}^n, & if \ m = n \\ 0, & if \ m \neq n \end{cases}, \tag{12}$$

in which $\boldsymbol{h}^n = \begin{bmatrix} \alpha_{21}^n & \alpha_{31}^n & \cdots & \alpha_{K1}^n \end{bmatrix}$. Note that if $P_k^n = 0$ and $\lambda_k^n > 0$, all the derivations above still apply by removing the $k$ th column and $k$ th row from $\boldsymbol{G}, NE_n\left(\boldsymbol{P_1}\right), \boldsymbol{g}^n, \boldsymbol{\nu}$ correspondingly. Hence, we can conclude $\frac{\partial I_1^n}{\partial P_1^n}$ is a constant $c\left(\boldsymbol{\alpha}^n, sign\left(NE_n\left(\boldsymbol{P_1}\right)\right)\right)$ that depends on both $\boldsymbol{\alpha}^n$ and the non-zero elements of $NE_n\left(\boldsymbol{P_1}\right)$.

*2)* If there exists $k \in \{2, \cdots, K\}$ satisfying $P_k^n = 0$ and $\lambda_k^n = 0$, the stationary interference caused to user $k$ is the same as its water-level $\nu_k$. Therefore, a sufficiently small increment or decrement $\varepsilon$ in user 1's allocated power $P_1^n$ may cause $sign\left(NE_n\left(\boldsymbol{P_1}^{n+}\right)\right)$ and $sign\left(NE_n\left(\boldsymbol{P_1}^{n-}\right)\right)$ to be different, i.e. the stationary interference $NE_n\left(\boldsymbol{P_1}\right)$ lies on the boundary between two polyhedra that have different piece-wise affine water-filling functions [3]. We need to treat the left-sided and the right-sided first derivatives respectively, and similar conclusions can be derived in the same way as in the first part. ∎

## Appendix B

*Proof of Proposition 2*

To solve the CE, the optimization solving CE in Table I is essentially

$$\max_{\{P_1^n\}} \sum_{n=1}^{N} \log_2\left(1 + \frac{P_1^n}{\sigma_1^n + \beta^n - \gamma^n P_1^n}\right)$$
$$s.t. \ \ P_1^n \geq 0, \beta^n - \gamma^n P_1^n \geq 0 \ and \ \ \sum_{n=1}^{N} P_1^n \leq \mathbf{P_1^{max}}. \tag{13}$$

Define $f_{a_1,a_2,b}(x) = \ln\left(1 + \dfrac{x}{a_1 + a_2 - bx}\right)$ in which the term $a_1 \geq 0$ represents the noise PSD. The second derivative of $f_{a_1,a_2,b}(x)$ is $f''_{a_1,a_2,b}(x) = -\dfrac{(a_1+a_2)\left[-(a_1+a_2)(2b-1) + 2b(b-1)x\right]}{(a_1+a_2-bx)^2\left[a_1+a_2-(b-1)x\right]^2}$ . Clearly, if $b \neq 0$ ,

$f''_{a_1,a_2,b}(x)$ is not always negative. We restrict the domain of $f_{a_1,a_2,b}$ to be

$$\boldsymbol{dom}\, f_{a_1,a_2,b} = \{x \geq 0\} \cap \{a_2 - bx \geq 0\},$$

because $x$ is the transmitted power and $a_2 - bx$ represents the stationary interference, both of which are non-negative. We derive a sufficient condition that guarantees $f_{a_1,a_2,b}(x)$ is concave in $\boldsymbol{dom}\, f_{a_1,a_2,b}$ :

$$a_2 > 0 \;\; and \;\; b < 0.5\left(1 - \frac{a_2}{a_1}\right). \tag{14}$$

This condition can be simply verified by using inequality analysis. Clearly, $a_2 > 0$ leads to $a_1 + a_2 > 0$

and $b < 0.5\left(1 - \dfrac{a_2}{a_1}\right) < 1$ . Therefore, $f''_{a_1,a_2,b}(x) < 0$ is equivalent to $-(a_1+a_2)(2b-1) + 2b(b-1)x > 0$ . We

have $x \in \boldsymbol{dom}\, f_{a_1,a_2,b} \Rightarrow a_2 - bx \geq 0 \Rightarrow bx - a_2 \cdot \dfrac{a_1+a_2}{a_2} \cdot \dfrac{b-0.5}{b-1} < 0 \Rightarrow -(a_1+a_2)(2b-1) + 2b(b-1)x > 0$ ,

because $\dfrac{a_1+a_2}{a_2} \cdot \dfrac{b-0.5}{b-1} > 1$ when $b < 0.5\left(1 - \dfrac{a_2}{a_1}\right)$ ,. Hence, the condition (14) leads to $f''_{a_1,a_2,b}(x) < 0$ .

Based on sufficient condition (14), we can see, if $\beta^n > 0 \;\; and \;\; \gamma^n < 0.5\left(1 - \dfrac{\beta^n}{\sigma_1^n}\right)$ for any $n \in \{1,\cdots,N\}$ ,

problem (13) belongs to convex programming. Therefore, if the following sufficient condition

$$\textbf{SC 1}: \left. \left(I_1^n - P_1^n \cdot \frac{\partial I_1^n}{\partial P_1^n}\right)\right|_{P_1 = P_{SE}} > 0 \;\; and \;\; -\left.\frac{\partial I_1^n}{\partial P_1^n}\right|_{P_1 = P_{SE}} < \frac{1}{2} - \frac{1}{2\sigma_1^n} \cdot \left.\left(I_1^n - P_1^n \cdot \frac{\partial I_1^n}{\partial P_1^n}\right)\right|_{P_1 = P_{SE}}$$

holds, SE satisfies the KKT optimality condition and solves the convex programming problem (13), i.e. SE

is also a CE. ∎

## Appendix C

*Proof of Proposition 3*

If the water-filling game $\mathcal{G}$ satisfies condition $\mathsf{SC\,1}$, then problem (13) is convex. We can use the following "maximum theorem" [3][18, p.116] to show that $P_1\,(\beta,\gamma)$ is continuous.

(*Maximum theorem*) Let $\phi(x,y)$ be a real-valued continuous function with domain $X \times Y$, where $X \subset \mathcal{R}^m$ and $Y \subset \mathcal{R}^n$ are closed and bounded sets. Suppose that $\phi(x,y)$ is strictly concave in $x$ for each $y$. The functions $\Phi(y) = \arg\max\{\phi(x,y) : x \in X\}$ is well-defined for all $y \in Y$, and is continuous.

We can restrict the domain of parameters $\beta$ and $\gamma$ in closed and bounded set, e.g. $\left|\gamma^n\right| \le M^+$, in which $M^+$ is a bound satisfying $M^+ \gg \max_n \boldsymbol{h}^n \left(\boldsymbol{I} + \boldsymbol{G}\right)^{-1} \boldsymbol{g}^n$. Apply the maximum theorem with $\phi = R\left(\boldsymbol{P_1},\beta,\gamma\right) = \sum_{n=1}^{N} f_{\sigma_1^n,\beta^n,\gamma^n}\left(P_1^n\right)$. The optimal solution $P_1\,(\beta,\gamma)$ of problem (13) is the function $\Phi$ in the maximum theorem, and hence is a continuous function of $(\beta,\gamma)$. As a result, $F(\beta,\gamma)$ is also continuous in $(\beta,\gamma)$. Note that $P_1\,(\beta,\gamma)$ and $F(\beta,\gamma)$ are not necessarily continuously differentiable.

By the definition of $F(\beta,\gamma)$ and conjectural equilibrium, we have that $F(\beta,\gamma) = \boldsymbol{0}$ implies conjectural equilibrium. Note $F(\beta_*,\gamma_*) = \boldsymbol{0}$. If there exist open neighborhoods $A \subset \mathcal{R}^N$ and $B \subset \mathcal{R}^N$ of $\beta_*$ and $\gamma_*$, and for $\forall \gamma \in B$, $F(\bullet,\gamma) : A \to \mathcal{R}^N$ is locally one-to-one, by the implicit function theorem [19], there exists open neighborhoods $A_0 \subset \mathcal{R}^N$ and $B_0 \subset \mathcal{R}^N$ of $\beta_*$ and $\gamma_*$ such that for each $\gamma \in B_0$, there is a unique $\beta(\gamma)$ satisfying $F\left(\beta(\gamma),\gamma\right) = \boldsymbol{0}$. Therefore, $\mathcal{G}$ admits an infinite set of conjectural equilibria.

Alternatively, we can view $F(\beta,\gamma) = \boldsymbol{0}$ as $N$ equations with $2N$ unknowns, hence, the equilibrium is usually not a single point but a continuous surface. We can explore the structure of $P_1\,(\beta,\gamma)$ to derive the expression of this surface. Particularly, under condition $\mathsf{SC\,1}$, the solution of convex problem (13) satisfies

$$\gamma^n\left(\gamma^n - 1\right)\left(P_1^n\right)^2 - \left(\sigma_1^n + \beta^n\right)\left(2\gamma^n - 1\right)P_1^n + \left(\sigma_1^n + \beta^n\right)^2 - \frac{\sigma_1^n + \beta^n}{\mu_1 - \lambda_1^n} = 0, \tag{15}$$

where $\lambda_1^n$ and $\mu_1$ are the Lagrange multipliers as in (13). The optimal $\boldsymbol{P}_1(\boldsymbol{\beta}, \boldsymbol{\gamma}) = \{P_1^n(\boldsymbol{\beta}, \boldsymbol{\gamma})\}$ is given by

$$P_1^n(\boldsymbol{\beta}, \boldsymbol{\gamma}) = \frac{(\sigma_1^n + \beta^n)(2\gamma^n - 1) + \sqrt{(\sigma_1^n + \beta^n)^2 (2\gamma^n - 1)^2 - 4\gamma^n (\gamma^n - 1)\left[(\sigma_1^n + \beta^n)^2 - \frac{\sigma_1^n + \beta^n}{\mu_1 - \lambda_1^n}\right]}}{2\gamma^n (\gamma^n - 1)}. \quad (16)$$

Note that the other root of equation (15) is removed by checking its feasibility in $\boldsymbol{dom}\, f_{\sigma_1^n, \beta^n, \gamma^n}$. By substituting (16) into (11) and (9), we can explicitly express $F(\boldsymbol{\beta}, \boldsymbol{\gamma})$ in terms of $\boldsymbol{\beta}$ and $\boldsymbol{\gamma}$, resulting in a very complex form of the surface. ∎